\documentclass{article}
\usepackage{graphicx}
\usepackage{epsfig}
\usepackage{amsfonts}
\usepackage{amssymb}

\newtheorem{theorem}{Theorem}
\newtheorem{definition}[theorem]{Definition}
\newtheorem{lemma}[theorem]{Lemma}
\newtheorem{proposition}[theorem]{Proposition}
\newtheorem{remark}[theorem]{Remark}
\def\erf{\mathop\mathrm{erf}\nolimits}
\def\div{\mathop\mathrm{div}\nolimits}
\def\Tr{\mathop\mathrm{Tr}\nolimits}
\def\spec{\mathop\mathrm{spec}\nolimits}

\title{Time of Arrival from Bohmian Flow}

\author{Gebhard Gr\"{u}bl and Klaus 
Rheinberger\thanks{Present address: Institut f\"{u}r 
An\"{a}stesiologie und Intensivmedizin, Universit\"{a}t Innsbruck, 
Anichstr. 35, A-6020 Innsbruck, Austria}\\[10pt] Institut f\"{u}r Theoretische
Physik der Universit\"{a}t
 Innsbruck\\
Technikerstr. 25\\A-6020 Innsbruck,
Austria\\[5pt]E-mail: gebhard.gruebl@uibk.ac.at}

\date{}

\begin{document}

\maketitle

\begin{abstract}
We develop a new conception for the quantum mechanical arrival time
distribution from the perspective of Bohmian mechanics. A detection
probability for detectors sensitive to quite arbitrary spacetime domains is
formulated. Basic positivity and monotonicity properties are established. We show
that our detection probability improves and generalises an earlier proposal by
Leavens and McKinnon. The difference between the two notions is illustrated
through application to a free wave packet.
\end{abstract}




\newpage
\section{Introduction}

According to quantum theory, the probability of the macroscopic event, which
is caused by a microscopic system during an act of measurement, is of the type
$\Tr(\rho_{t_{0}}E)$. Here $E$ is an orthogonal projection within the system's
Hilbert space and $\rho_{t_{0}}$ is the system's density operator at time
$t_{0}$. The time $t_{0}$, at which the approximately instantaneous
measurement interaction takes place, is determined by the experiment's design.
There are, however, important situations which do not - even approximately -
fit into this framework in any obvious way. Consider for instance an unstable
nucleus, which is monitored for several days by some initially activated
detector. You patiently sit next to the detector and register the time, when
you hear the click. What is the probability that you hear the click during a
certain time interval? A cleaner model situation involves a freely propagating
one particle wave packet, which slowly sweeps over a detector activated at
time $0$. The detector is small compared to the wave packet's size. What is
the probability $P(T)$ of a click, happening at any time $t$ in the range
$0<t<T$? The funcion $P$ is called \textit{arrival time distribution}. Several
proposals try to answer this question without reproducing the quantum Zeno
paradox \cite{MS}. Let us describe them briefly. For an extensive summary
of the subject see \cite{ML}.

A first approach attempts to fit $P(T)$ into the scheme $\Tr(\rho_{t_{0}}E)$
through quantising the phase space function $A$, which represents the
classical time of arrival. The function $A$ maps each phase space point from
its domain onto that finite time, at which (according to the system's
classical dynamics) this point enters the detector's location. With $E$ being
the spectral projection of the quantised $A$ associated with the spectral
interval $(0,T)$ it is assumed that $P(T)=\Tr(\rho_{0}E)$. Working out this
general idea reveals that ad hoc regularisation assumptions are needed, in
order to obtain a self adjoint quantisation of $A$ \cite{GRT}. The need for
regularisation is both due to the unboundedness of $A$ around $p=0$ and to a
classically unspecified operator ordering. Due to its regularisation
ambiguity, this definition of $P(T)$ does not seem convincing.

A second strategy attempts to derive the arrival time distribution from a
unitary quantum dynamical model of the continuing observation process and a
single final measurement, i.e. the ``reading out of the observer's notices''.
To this end, an auxilliary quantum system is coupled to the particle during
the time interval $(0,T)$. The auxilliary system's position is taken as the
pointer position of a clock and its evolution is stopped through an
interaction with the particle's wave function \cite{AOP}. This approach does
not yield the picture of a sudden click happening at a certain time
$t\in(0,T)$, but rather of a smooth influence being exerted onto a position
distribution. Only the final observation at a controllable time $t_{0}>T$ then
produces the stochastic position value, which is interpreted as an approximate
time of arrival. Therefore, our macroscopic impression that facts are
permanently established in the course of time, instead of being created with a
final measurement only, - ``at the end of the day'', as Sheldon Goldstein has
phrased it\footnote{private communication}, - remains unexplained.

A third way of defining $P$ is obtained by exposing the particle's wave
function to an absorbing detector, whose influence onto the wave function is
modelled by a nonhermitean Hamiltonian, e.g. \cite{MPL}. Thereby the one
particle
 dynamics becomes nonunitary and the quantity $-\frac{d\left\|  \psi
_{t}\right\|  ^{2}}{dt}\left|  dt\right|  $ for $\left\|  \psi_{0}\right\|
=1$ is then - up to an overall normalisation - interpreted as the probability
density of clicks at time $t$. However in general, though not
in \cite{MPL}, $-\frac{d\left\|  \psi_{t}\right\|
 ^{2}}{dt}$ may take
negative values, which in turn implies that the
 probability
$P(T)=1-\left\|  \psi_{T}\right\|  ^{2}$ (for the detector to
 click sometimes
between time $0$ and time $T$) may decrease upon increasing
 $T$. Clearly, a
decreasing probability is questionable if one imagines say
 $1000$ independent
copies of the system side by side and the percentage of
 counters having made
their click is monitored as a function of time. If this
 percentage decreases
with $T$, a mechanism seems to be at work, which makes
 clicks unhappened!

Finally, Leavens \cite{L1993}, \cite{L1998} and McKinnon and Leavens
\cite{McL} have defined an arrival time distribution $P$, which is motivated
by the \textit{Bohmian flow} connected with a solution $\psi_{t}$ of
Schr\"{o}dinger's equation. They considered the one dimensional case and
argued that the (conditional) probability density of clicks equals
$const\cdot\left|  j(t,L)dt\right|  $, where $j(t,L)$ is the spatial
probability current density at the detector's location $L$ at time $t$. In
case of $\int_{0}^{\infty}dt\left|  j(t,L)\right|  =:c<\infty$, the
conditional arrival time distribution%
\begin{equation}
P(T)=\frac{1}{c}\cdot\int_{0}^{T}dt\left|  j(t,L)\right|
\label{stromdichtebetr}
\end{equation}
is a nondecreasing (nonnegative) function on the interval $\left(
0,\infty\right)  $. Yet the integral $\int_{0}^{\infty}dt\left|
j(t,L)\right|  $ need not be finite, as e.g. in the case of an harmonic
oscillator dynamics, where the mapping $t\mapsto j(t,L)$ is periodic. In such
cases therefore, the definition (\ref{stromdichtebetr}) does not make sense.

The probability density $\sim\left|  j(t,L)\right|  $ is derived by Leavens as
the ``infinitesimal'' probability that the particle's Bohmian trajectory
passes the point $L$ during $dt$ at time $t$ provided the Bohmian position at
time $0$ is distributed by $\left|  \psi_{0}\right|  ^{2}\left|  dx\right|  $.
If one assumes that the detector clicks each time it intersects with the
particle's Bohmian trajectory, the density $\left|  j(t,L)dt\right|  /c$
indeed yields the probability density of clicks. This seems to be a reasonable
idealisation if the detector is active during a short time interval. What
happens, however, if the detector is active over a longer period of time, such
that the same trajectories pass the detector more than once? Does a detector
really increase its click probability when trajectories intersect, which have
done so before?

The possibility of multiple intersections between Bohmian trajectories and
detector positions has already been taken into account in \cite{DDGZ},
\cite{TDM}. These works have ruled out multiple intersections for scattering
situations. For a summary see also chapter 16 of D\"{u}rr's recent text book
\cite{DBuch} or \cite{CFG}. As a next step, in the context of ``near flield scattering'' the exit
time statistics from a large but finite sphere around the scattering center has
been investigated in \cite{DDGZ2}. In case of multiple crossings of the sphere's
surface by Bohmian trajectories, replacement of $j$ in equation (\ref{stromdichtebetr})
by a truncated current has been proposed in \cite{DDGZ2}, in order to obtain the
correct exit time statistics. The truncated current only counts the first exit
of trajectories as detection events. This is reasonable if the initial wave packet is well
localised within the sphere. Yet if a considerable part of the wave packet has left the sphere
by the time the detector is activated, those trajectories,
which have already entered the detector and stay there, carry a
nonnegligible portion of probability. Accordingly they should contribute to the detector's click probability.
However these trajectories do not contribute to the surface integral of the truncated current. Thus under such circumstances
a more general prescription is needed to count also those trajectories which are confined to the detection
volume during the full period of detector activity.

In this work we propose and explore a very natural
definition of detection probability within Bohmian mechanics, which on the one hand meets
the above needs for generalisation but also implies the idea
of using the truncated current for the exit time problem described in \cite{DDGZ2}. The
physical argument behind it is quite simple: a realised trajectory induces a detection event
at the earliest instance only, when this trajectory falls into the detector's volume
because thereafter the detector remains discharged. According to Bohmian mechanics
each individual trajectory is realised with an "infinitesimal" probability to be computed from the wave function.
Adding up these infinitesimal probabilities for all the
trajectories intersecting the detector's volume during its period of activity
then yields this detector's
click probability. Assuming this, we obtain an expression for the arrival time
probability density, which in general depends on
the spacetime region to which the detector is sensitive. In cases, where each Bohmian
trajectory crosses a (point like) detector at most once, equation
(\ref{stromdichtebetr}) remains valid. However, our definition yields
different probabilities otherwise.

Why an experimental decision between the various conflicting proposals for
$P(T)$ is not yet feasible has been indicated in section 10 of the latest
review of the subject \cite{ML}. The basic reason seems to lie in the difficulties in
preparing a specific wave packet which has to be large compared to
the detector's size and which in addition has to pass the detector
sufficiently slowly. Clearly such experiments are not precluded
on principle.

After a brief summary of Bohmian mechanics in section 2, we develop our
definition of the arrival time distribution $P(T)$ in sections 3, 4, and 5
within a Galilean spacetime frame work. Using spacetime proves very
suggestive since Bohmian trajectories become one dimensional submanifolds
(worldlines) instead of mappings. In section 3 we define the Galilean (one
particle) spacetime from its structural atlas. Section 4 contains an outline
of conserved flows on Galilean spacetime. Here again we choose the coordinate
independent formalism of exterior calculus. This has the following advantage.
When computing the flux through a (possibly moving) hypersurface it is the current 3-form
which is integrated over a 3-manifold. Neither a metric spacetime structure
nor a normal vector field, both breaking Galilean invariance, need to be
introduced. The main result of this section is the formula of definition 4
which gives the amount of conserved ''mass'' passing through a spacetime
region $X$. In section 5 we apply this formula to the flow of the quantum
mechanical position probability.
Here our definition of the detection probability is obtained from the quantum
mechanical probability measure on the set of Bohmian worldlines (orbit
space), which follows from the wave function $\psi$ under consideration. The
probability that a detector clicks, is assumed to equal the probability
measure of the set $\widetilde{X}$ of all those orbits, that have a nonempty
intersection with the
spacetime region $X$, to which the detector is sensitive. This measure in
turn equals the usual quantum mechanical probability measure to detect a
particle with wave function $\psi$ at $t=0$ within the set of all those
locations which are taken by the orbits of $\widetilde{X}$ at $t=0$. Our
definition works for very general, extended spacetime regions and it works
for the free Schr\"{o}dinger dynamics as well as for ones with nonzero
potential. The coordinate independent treatment guarantees Galilean invariance
of the detection probability for zero potential. In section 6 we illustrate our
notion of $P(T)$ through the example of a free standing Gaussian wave
packet.

Our definition of $P(T)$ in terms of the Bohmian flow could be
improved by taking into account the detector's influence onto the particle's
Bohmian trajectories. Since the latter become projections of the higher
dimensional orbits of the detector plus particle system, this effect can be
considerable even for detectors without any back reaction onto the particle
wave function \cite{AV}. The general idea of our approach however, remains the
same. Also if the detector (or even the observer) is modeled as part of the
quantum system, an assumption has to be made about when each individual orbit
generates the click (in the observer's mind). This rule then mathematically
represents the discrete event, which is missing from standard quantum theory.

\section{Summary of Bohmian Mechanics}

The density operator $\rho$, representing the state of a quantum system with (separable) Hilbert
space $\mathcal{H}$, defines a probability measure $W_{\rho,A}$ on the spectrum
of any self adjoint Operator $A$ of $\mathcal{H}$. It is given by $W_{\rho
,A}(I)=\Tr(\rho E_{A}(I))$ where $E_{A}(I)$ denotes the spectral projection of
$A$ associated with the Borel set $I\subset \spec(A)\subset\mathbb{R}$.
Standard quantum theory assumes that, if a measurement of the observable $A$
is performed on the state $\rho$, then $W_{\rho,A}(I)$ equals the probability
of the event ''the measured spectral value $a$ of $A$ belongs to $I$''. Now,
for $\dim(\mathcal{H)}\geq2$, there does not exist a density operator $\rho$
such that $W_{\rho,A}$ is dispersion free, i.e. a point measure, for all $A$.
Gleason \cite{Gle} has investigated the
question whether there exist more general ways of defining a probability
measure on $\spec(A)$ for all $A$. To this end he considered the mappings
$\sigma$ from the set $\Pi$ of all orthogonal projections of $\mathcal{H}$
into the real numbers such that $\sigma(P)\geq0$ for all $P\in\Pi$ and
$\sigma(\sum_{i}P_{i})=\sum_{i}\sigma(P_{i})$ for any countable sum of
$P_{i}\in\Pi$, with $P_{i}P_{j}=\delta_{i,j}P_{i}$. In addition he assumed
$\sigma(id)=1$. From this he derived in case of $\dim(\mathcal{H)}\geq3$ that for any such mapping $\sigma
:\Pi\rightarrow\left[  0,1\right]  $ there exists a density operator $\rho$
such that $\sigma(P)=\Tr(\rho P)$ for all $P\in\Pi$. Thus the idea of
generalising the formula $W_{\rho,A}(I)=\Tr(\rho E_{A}(I))$ to $W_{\sigma
,A}(I)=\sigma(E_{A}(I))$ in order to possibly obtain ''deterministic states'',
i.e. point measures $W_{\sigma,A}$ for all $A$, and a representation of density operators as
mixtures of these, under the adopted assumptions fails.

The standard quantum physical interpretation of this body of mathematical facts
leads to the following conclusion.
It is inconsistent to suppose that the state of an individual quantum system
is a deterministic state, i.e.~determines values for all observables, and
it is inconsistent to suppose that a density operator $\rho$ only
describes a mixture of such fictitious deterministic states.
(It is generally held inconsistent to suppose that an idividual particle has
a specific position and a specific momentum and so on.)

From this conclusion
then the notorious \textit{quantum measurement problem} follows: How can standard quantum theory
represent within its formalism the mere fact that individual closed
systems do have properties? (This
surely is the case for systems comprising an observer and not beeing in need of
any sort of external observation inducing a state reduction, the quantised deus ex machina.)
Which fact concerning a closed system is it,
whose probability of being the case is given by $W_{\rho,A}(I)$ ?

Bohmian mechanics resolves these problems for systems with a
Schr\"{o}dinger (or Dirac) equation: a picture of individual
systems with defined properties emerges. A concise review of Bohmian mechanics is given in
reference \cite{Be}. An informal but clear
summary is to be found in Goldstein's contribution to the Stanford
Encyclopedia of Philosophy \cite{Go}. Let us summarise the basic ideas.

Bohmian mechanics introduces deterministic
states which violate Gleason's assumptions and accordingly circumvent his theorem. It is assumed that
an individual system has a state $(\psi,q)$ given by a wave
function $\psi$ in the system's Hilbert space and a point $q$ in its
configuration space. $q$ is supposed to represent the actual positions of the
system's constituents. Other observable properties of the system have to
be derived from the Bohmian state through
a dynamical analysis of the concrete experiment
designed to measure them. In this way all other properties like spin or
momentum are expressed through the state's well defined position properties. It turns out, that the spectral value,
which a general observable assumes in a Bohmian state, depends on the specific way of
how this observable is measured, i.e. contextuality is found to be realised \cite{Ha}.
Accordingly Gleason's assumptions on the mapping $\sigma$ are violated because $\sigma$ needs a much more complex domain
than simply the set of all orthogonal projections.

In order to work out the dynamical program of reducing all state properties to position
properties, the time evolution of Bohmian states
is needed. It is assumed to be given by a Schr\"{o}dinger equation
for the wave function and by a time dependent tangent vector field $v$ on the
configuration space. $v$ is defined in terms of the solution $\psi_{t}$ of
the adopted Schr\"{o}dinger equation with initial condition $\psi_{0}=\psi$. The
integral curve $\gamma_{q}$ of $v$ with initial condition $\gamma_{q}(0)=q$
gives the system's configuration at time $t$ by $\gamma_{q}(t)$.

Finally, Bohmian mechanics establishes contact with empirical data. This
happens according to the rule of quantum equilibrium. It states that for an
ensemble of systems, each with wave function $\psi$, the individual system's
position $q$ belongs to a configuration space domain $\Delta$ with the usual
probability $\int_{\Delta}\left|  \psi\right|  ^{2}d^{n}q$. A controlled
preparation of $q$ contradicting this rule is assumed to be impossible by
present day technology. (All this can be justified to a certain extent within
the Bohmian picture \cite{Be}.) And finally as a last ingredient it is supposed that it is the
center of mass position of pointers and the like that we observe.

The rules of Bohmian mechanics are such that the probabilistic statements of
standard quantum theory are reproduced. So there seems no room left to argue
about the empirical superiority of either standard quantum mechanics or its
Bohmian extension at the ensemble level. Bohmian mechanics might, however,
give a clue for the correct treatment of ensemble problems, where the standard
interpretation remains unclear and offers conflicting strategies. As described above, standard quantum theory
offers various different conceptions for the arrival time distribution $P(T)$. Therefore we hold the arrival time problem
to be one such opportunity for Bohmian mechanics to possibly show that it also
has its value in dealing with ensemble problems on top of its merit of providing a
language for speaking about individual systems.
We add another conception for $P(T)$ which is motivated by the
Bohmian extension of quantum mechanics. It does not conform to the standard
scheme of identifying $P(T)$ with some quantity of the type $\Tr(\rho E)$ with $E$ being independent from $\rho$. Our
$P(T)$ needs the concept of Bohmian trajectories for its very formulation. One
should note, however, that Bohmian trajectories are implicit in the wave
function, whether one intends to make use of them or not.

\section{Galilean Spacetime}

We model spacetime as a \textit{Galilean manifold}. Various equivalent
definitions of a Galilean manifold can be given. Here we use the method of a
structural atlas. The basic object is the group $\mathcal{G}$ of
\textit{(orthochronous) Galilei transformations}.
\begin{eqnarray}
\Gamma & := & \left\{  \left( {1\atop v}{0\atop R}
\right)  \in Gl_{n+1}(\mathbb{R})\mid v\in\mathbb{R}^{n},R\in O_{n}\right\}
,\label{Galgr}\\
\mathcal{G}  & := & \left\{  g:\mathbb{R}^{n+1}\rightarrow\mathbb{R}^{n+1}
,g(\xi)=\gamma\cdot\xi+a\mid\gamma\in\Gamma,a\in\mathbb{R}^{n+1}\right\}
.\nonumber
\end{eqnarray}
The elements of $\mathbb{R}^{n+1}$ and $\mathbb{R}^{n}$ are treated as column
vectors throughout the text.

\begin{definition}
A Galilean manifold $(\mathcal{M},\mathcal{A}_{\mathcal{G}})$ consists of a
differentiable manifold $\mathcal{M}$ and a subset $\mathcal{A}_{\mathcal{G}}
$ of the atlas $\mathcal{A}$ of $\mathcal{M}$, where $\mathcal{A}
_{\mathcal{G}}$ contains global charts only and the set of transition functions
$\left\{  \Phi_{2}\circ\Phi_{1}^{-1}\mid\Phi_{1},\Phi_{2}\in\mathcal{A}
_{\mathcal{G}}\right\}  $ equals $\mathcal{G}$. The charts $\Phi\in
\mathcal{A}_{\mathcal{G}}$ are called Galilean charts.
\end{definition}

A Galilean manifold carries the canonical time-1-form $\theta:=d\Phi^{0}$ with
$\Phi=(\Phi^{0},\Phi^{1},..\Phi^{n})^{t}\in\mathcal{A}_{\mathcal{G}}$. Observe
that $\theta$ is independent from the choice of $\Phi$. Tangent vectors $v\in
T(\mathcal{M})$ with $\theta(v)=1$ are called \textit{velocity vectors}, and
tangent vectors with $\theta(v)=0$ are called \textit{spacelike vectors}. The
subbundle $\mathcal{R}(\mathcal{M}):=\ker(\theta)$ of spacelike vectors is
completely integrable. Its integral manifolds are given by $\Sigma_{\Phi
,t}:=\left\{  p\in\mathcal{M}\mid\Phi^{0}(p)=t\right\}  $ where $\Phi
\in\mathcal{A}_{\mathcal{G}}$ and $t\in\mathbb{R}$. These integral manifolds
are called \textit{instantaneous spaces}.

The vector bundle $\mathcal{R}(\mathcal{M})$ carries a canonical positive
definite fibre metric
\[
\left\langle \cdot,\cdot\right\rangle :=\sum_{k=1}^{n}d\Phi^{k}\otimes
d\Phi^{k},
\]
where the restriction of $d\Phi^{k}$ to $\mathcal{R}(\mathcal{M})$ is again
denoted as $d\Phi^{k}$. Note that $\left\langle \cdot,\cdot\right\rangle $ is
well defined as a fibre metric of $\mathcal{R}(\mathcal{M})$, but is not so as
a fibre metric of $T(\mathcal{M})$. Finally, the Galilean manifold carries two
orientations represented by the two volume $(n+1)$-forms
\[
\Omega:=\left\{  \pm d\Phi^{0}\wedge d\Phi^{1}\wedge...\wedge d\Phi
^{n}\right\}  ,\Phi\in\mathcal{A}_{\mathcal{G}}\mbox{.}
\]
Thus the density $\left|  d\Phi^{0}\wedge d\Phi^{1}\wedge...\wedge d\Phi
^{n}\right|  $ is unique. Various further structures are canonically defined
on $(\mathcal{M},\mathcal{A}_{\mathcal{G}})$ as e.g. a linear connection of
the tangent bundle. We shall not use them here.

\section{Conserved Flows}

Let $j$ be a differentiable tangent vector field on a Galilean manifold
$(\mathcal{M},\mathcal{A}_{\mathcal{G}})$. By choosing a volume form
$\omega\in\Omega$ the differentiable $n$-form $J$ on $M$ is obtained through
\[
J:=j\lrcorner\omega:\left(  t_{1},..t_{n}\right)  \mapsto\omega(j,t_{1}%
,..t_{n}).
\]
The associated density $\left|  J\right|  $ does not depend on the chosen
$\omega$. The \textit{divergence} of $j$ is the unique function (see e.g. page
281 of \cite{Kob}) $\div(j)$ satisfying
\begin{equation}
L_{j}\omega=\div(j)\omega.\label{div}
\end{equation}
Here $L_{j}\omega$ denotes the Lie derivative of $\omega$ with respect to $j$.
The divergence of $j$ does not depend on the choice $\omega\in\Omega$. Observe
that this definition of the divergence of a vector field does not make use of
a (pseudo-) Riemannian metric. It is built on a given density $\left|
\omega\right|  $. As it is the case with a Galilean manifold, this density
$\left|  \omega\right|  $ need not be induced by a (pseudo-) Riemannian
metric. If $\left|  \omega\right|  $ is the metric density of a (pseudo-)
Riemannian manifold, the above definition for $\div(j)$ coincides with the
usual one.

There holds $L_{j}\omega=j\lrcorner d\omega+d(j\lrcorner\omega)=dJ$, and
therefore
\[
\div(j)=0\Leftrightarrow dJ=0.
\]
Furthermore, if $\div(j)=0$, then $L_{j}J=j\lrcorner dJ+d(j\lrcorner
J)=j\lrcorner dJ=0$. In terms of a Galilean chart $\Phi$ the divergence of the
vector field $j=\sum_{k=0}^{n}j_{\Phi}^{k}\cdot\partial_{k}^{\Phi}$ reads
\[
\div(j)=\sum_{k=0}^{n}\partial_{k}^{\Phi}\left[  j_{\Phi}^{k}\right]  .
\]
Here $\partial_{k}^{\Phi}$ denotes the tangent vector field $\frac{\partial
}{\partial\Phi^{k}}$ associated with the chart $\Phi$ and $j_{\Phi}^{k}$ are
the coefficient functions of $j$ with respect to the coordinate frame
$\underline{\partial}^{\Phi}:=\left(  \partial_{0}^{\Phi},...\partial
_{n}^{\Phi}\right)  $. For $\omega=d\Phi^{0}\wedge d\Phi^{1}\wedge...\wedge
d\Phi^{n}$ the $n$-form $J$ is given by
\[
J=\sum_{k=0}^{n}(-1)^{k}\cdot j_{\Phi}^{k}\cdot d\Phi^{0}\wedge...\wedge
d\Phi^{k-1}\wedge d\Phi^{k+1}...\wedge d\Phi^{n}.
\]

Let $j$ be a $\mathit{C}^{1}$-vector field on $\mathcal{M}$ such that
$\theta_{p}(j)\neq0$ for all $p\in\mathcal{M}$. Then the velocity vector field
of $j$ is defined on $\mathcal{M}$ by $\widehat{j}:=\frac{1}{\theta(j)}j$. The
maximal integral curve of $\widehat{j}$ through $p\in\mathcal{M}$ is the
(unique) function $\gamma:I\rightarrow\mathcal{M}$ with $\gamma(0)=p$ and
\[
\dot{\gamma}(\lambda)=\widehat{j}_{\gamma(\lambda)}\mbox{ for all }\lambda\in
I.
\]
Here $I$ is an open real interval, which cannot be extended. The image
$\gamma(I)\subset\mathcal{M}$ is called (integral) orbit of $\widehat{j}$
through $p$.

Assume the vector field $\widehat{j}$ on $\mathcal{M}$ to be complete, i.e.
each maximal integral curve of $\widehat{j}$ has the domain $\mathbb{R}$. Then
a unique one parameter group of mappings $F_{s}:\mathcal{M}\rightarrow
\mathcal{M}$ with $F_{s}(p)=\gamma(s)$ exists, where $\gamma$ is the maximal
integral curve of $\widehat{j}$ with $\gamma(0)=p$. There holds $F_{s}\circ
F_{t}=F_{s+t}$ and $F_{s}^{-1}=F_{-s}$ for all $s,t\in\mathbb{R}$. The mapping
$F:\mathbb{R}\times\mathcal{M}\rightarrow\mathcal{M},(s,p)\mapsto F_{s}(p)$ is
called the flow of $\widehat{j}$. Since $\theta(\widehat{j})=1$, for the
maximal integral curve through any $p\in\mathcal{M} $ there holds $\left(
\Phi^{0}\circ\gamma\right)  (s)=\Phi^{0}(p)+s$ for any $s\in\mathbb{R}$ and
for any $\Phi\in\mathcal{A}_{\mathcal{G}}$. Thus no orbit begins or ends at
finite time. In particular $F_{s}$ carries instantaneous spaces into
instantaneous spaces, i.e. $F_{s}(\Sigma_{\Phi,t})=\Sigma_{\Phi,t+s}$.

If now $\div(j)=0$, we have $dJ=0$. From this and because of $j\lrcorner
J=j\lrcorner(j\lrcorner\omega)=0$ there follows $L_{\widehat{j}}J=\widehat
{j}\lrcorner dJ+d(\frac{1}{\theta(j)}j\lrcorner J)=0$ and therefore both $J$
and $\left|  J\right|  $ are invariant under the pull back with the flow of
$\widehat{j}$, i.e. $F_{s}^{\ast}J=J$ and also $F_{s}^{\ast}\left|  J\right|
=\left|  J\right|  $ for all $s\in\mathbb{R}$. From the pull back formula for
integrals of differential forms then the following lemma follows.

\begin{lemma}
[Integral conservation law]Let $j$ be a $\mathit{C}^{1}$-vector field on
$\mathcal{M}$ such that $\div(j)=0,$ $\theta_{p}(j)\neq0$ for all
$p\in\mathcal{M}$ and such that $\widehat{j}$ is complete. $F$ denote the flow
of $\widehat{j}$. Then for any Borel set of an instantaneous space
$X\subset\Sigma_{\Phi,s}$ and for any $t\in\mathbb{R}$ there holds
\begin{equation}
\int_{F_{t}(X)}\left|  J\right|  =\int_{X}\left|  J\right|  \mbox{.}
\label{probTransp}
\end{equation}
\end{lemma}

\begin{remark}
Depending on the physical context an integral of the type $\int_{X}\left|
J\right|  $ is interpreted as the mass or probability ``contained'' in the
instantaneous region $X$. The lemma thus establishes the picture of a flow
which transports mass or probability without change along the flow lines. The
same amount of mass which is contained in an instantaneous region $X$ is
contained in $F_{t}(X)$ for any $t\in\mathbb{R}$.
\end{remark}

Consider now more general sets $X\subset\mathcal{M}$ which need not be
contained in an instantaneous subspace. Let us try to formulate a precise
notion of the \textit{amount of mass passing through }$X$. A clear and
unambiguous way of doing this is by determining the set $\widetilde{X}$ of all
orbits passing through $X$ and by computing the amount of mass carried by
these orbits. This can be done by intersecting these orbits with any
instantaneous space $\Sigma_{\Phi,t}$ and by integrating $\left|  J\right|  $
over this intersection. Thus we have motivated the following definition, which
is illustrated by figure 1.

\begin{definition}
Let $j$ be a $\mathit{C}^{1}$-vector field on $\mathcal{M}$ such that
$\div(j)=0,$ $\theta_{p}(j)\neq0$ for all $p\in\mathcal{M}$ and such that
$\widehat{j}$ is complete. $F$ denote the flow of $\widehat{j}$. Let $pr$ be
the projection $pr:\mathbb{R}\times\mathcal{M}\rightarrow\mathcal{M}
,(t,p)\mapsto p$ and let $E_{\Phi,t}$ be the restriction of the flow $F$ to
$\mathbb{R}\times\Sigma_{\Phi,t}$. Then $\pi_{\Phi,t}:=pr\circ E_{\Phi,t}
^{-1}$ is the fibre projection of $\mathcal{M}$ onto $\Sigma_{\Phi,t}$ along
the orbits of $\widehat{j}$ . If for a subset $X$ of $\mathcal{M}$ its
projection $\pi_{\Phi,t}(X)\subset\Sigma_{\Phi,t}$ is a Borel set, then we
define the transition $P\left[  X\right]  $ of $j$ through $X$ as
\[
P\left[  X\right]  :=\int_{\pi_{\Phi,t}\left(  X\right)  }\left|  J\right|
\in\left[  0,1\right]  .
\]
\end{definition}

\begin{remark}
Note that the transition $P\left[  X\right]  $ does not depend on the chosen
hypersurface $\Sigma_{\Phi,t}$. This follows immediately from $\pi_{\Phi
,s+t}=F_{t}\circ\pi_{\Phi,s}$ and from equation (\ref{probTransp}) because of
\[
\int_{\pi_{\Phi,s+t}\left(  X\right)  }\left|  J\right|  =\int_{F_{t}\left(
\pi_{\Phi,s}\left(  X\right)  \right)  }\left|  J\right|  =\int_{\pi_{\Phi
,s}\left(  X\right)  }\left|  J\right|  .
\]
\end{remark}

\begin{remark}
Let $X_{1},X_{2}\subset\mathcal{M}$ be disjoint. Then the sets $\pi_{\Phi
,t}(X_{1})$ and $\pi_{\Phi,t}(X_{2})$ need not be disjoint. As a consequence
$P\left[  X_{1}\cup X_{2}\right]  \neq P\left[  X_{1}\right]  +P\left[
X_{2}\right]  $ in general. Thus $P$ is not a measure$.$ Yet $X_{1}\subset
X_{2}$ implies $P\left[  X_{1}\right]  \leq P\left[  X_{2}\right]  $.
\end{remark}%

\begin{figure}[ht]
\centering
\includegraphics[height=2.2035in, width=2.0384in]{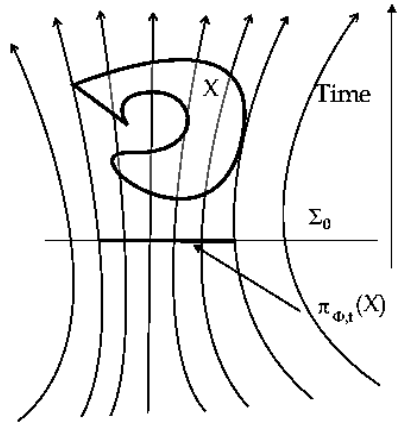}\\
\caption{transition of $j$ through $X$}
\end{figure}

%

\section{Detection probability from Bohmian flow}

In order to define a (free) Schr\"{o}dinger equation on a Galilean manifold
$(\mathcal{M},\mathcal{A}_{\mathcal{G}})$, one has to choose a tangent frame
$\underline{\partial}^{\Phi}$, which is associated with a Galilean chart
$\Phi$. Any two such \textit{Galilean charts} $\Phi_{1}$ and $\Phi_{2}$ are
connected by $\Phi_{2}=g\circ\Phi_{1}=\gamma\cdot\Phi_{1}+a$ with $\gamma
\in\Gamma$ and $a\in\mathbb{R}^{n+1}$. The frames then obey $\underline
{\partial}^{\Phi_{2}}=\underline{\partial}^{\Phi_{1}}\cdot\gamma$. In terms of
this matrix notation the duality between a frame and its co-frame
$d\Phi:=(d\Phi^{0},..d\Phi^{n})^{t}$ is expressed by the equation
$d\Phi(\underline{\partial}^{\Phi})=I_{n+1}$, with $I_{n+1}\in Gl_{n+1}
(\mathbb{R})$ being the unit matrix. There holds $d\Phi_{2}=\gamma^{-1}\cdot
d\Phi_{1}$. Note that $d\Phi_{2}=d\Phi_{1}$ and $\underline{\partial}
^{\Phi_{2}}=\underline{\partial}^{\Phi_{1}}$ for $\gamma=I_{n+1}$, such that a
chosen frame determines the chart $\Phi\in\mathcal{A}_{\mathcal{G}}$ up to an
element $a\in\mathbb{R}^{n+1}$.

For every Galilean frame $\underline{\partial}^{\Phi}$ we define the
differential operator $D_{\underline{\partial}^{\Phi}}$ operating on
$\mathit{C}^{2}$-functions $\psi:\mathcal{M}\rightarrow\mathbb{C}$ through
\[
D_{\underline{\partial}^{\Phi}}:=i\hbar\partial_{0}^{\Phi}+\frac{\hbar^{2}
}{2m}\sum_{k=1}^{n}\partial_{k}^{\Phi}\partial_{k}^{\Phi}.
\]
The operators $D_{\underline{\partial}^{\Phi}}$ depend on the frame
$\underline{\partial}^{\Phi}$ because of the term $i\hbar\partial_{0}^{\Phi}$.
If $\Phi_{1}$ and $\Phi_{2}$ are two \textit{Galilean frames} with
$\underline{\partial}^{\Phi_{2}}=\underline{\partial}^{\Phi_{1}}\cdot\gamma$
and $\gamma = \left({1\atop v}{0\atop R}\right)$
then there holds $\partial_{0}^{\Phi_{2}}=\partial_{0}^{\Phi_{1}
}+\sum_{k=1}^{n}v^{k}\partial_{k}^{\Phi_{1}}$. The following proposition
however, which can be checked easily, shows that the solution spaces
$\ker\left(  D_{\underline{\partial}^{\Phi}}\right)  $ can be mapped
bijectively onto each other.

\begin{proposition}
Let $\underline{\partial}^{\Phi_{1}},\underline{\partial}^{\Phi_{2}}$ be
Galilean frames with $\underline{\partial}^{\Phi_{2}}=\underline{\partial
}^{\Phi_{1}}\cdot\gamma$, and $\gamma=\left({1\atop v}{0\atop R}\right)$.
Let the function $\phi:\mathcal{M}\rightarrow\mathbb{R}$ be given
by
\[
\phi=\frac{m}{\hbar}\left(  \frac{v^{2}}{2}\Phi_{1}^{0}-\sum_{k=1}^{n}
v^{k}\Phi_{1}^{k}\right)  +c,\quad c\in\mathbb{R}.
\]
Then $\ker(D_{\underline{\partial}^{\Phi_{1}}})$ is mapped bijectively onto
$\ker(D_{\underline{\partial}^{\Phi_{2}}})$ through $\psi\mapsto\exp
(i\phi)\psi$.
\end{proposition}

Let $\psi$ solve the free Schr\"{o}dinger equation $D_{\underline{\partial
}^{\Phi}}\psi=0$. Then the \textit{current (vector field)} $j(\psi
,\underline{\partial}^{\Phi})$ is defined by
\begin{equation}
j(\psi,\underline{\partial}^{\Phi}):=\psi^{\ast}\psi\cdot\partial_{0}^{\Phi
}+\frac{\hbar}{2mi}\sum_{k=1}^{n}\left[  \psi^{\ast}\left(  \partial_{k}
^{\Phi}\psi\right)  -\psi\left(  \partial_{k}^{\Phi}\psi^{\ast}\right)
\right]  \cdot\partial_{k}^{\Phi}.\label{stromVF}
\end{equation}
Due to $D_{\underline{\partial}^{\Phi}}\psi=0$ there holds $\div\left(
j(\psi,\underline{\partial}^{\Phi})\right) =0$. The current's frame
independence follows through a straight forward computation.

\begin{proposition}
Let $\underline{\partial}^{\Phi_{1}},\underline{\partial}^{\Phi_{2}}$ be
Galilean frames and let $\psi\in\ker(D_{\underline{\partial}^{\Phi_{1}}})$.
Then $j(\psi,\underline{\partial}^{\Phi_{1}})=j(\exp(i\phi)\psi,\underline
{\partial}^{\Phi_{2}})$.
\end{proposition}

\begin{remark}
For $\psi\in\ker(D_{\underline{\partial}^{\Phi}})$ we thus abbreviate
$j:=j(\psi,\underline{\partial}^{\Phi})$.
\end{remark}

For $\psi\in\ker(D_{\underline{\partial}^{\Phi}})$ the unitarity of the
Schr\"{o}dinger evolution implies that the integral $\int_{\Sigma_{\Phi,t}
}\left(  \psi^{\ast}\psi\right)  \cdot\left|  d\Phi^{1}\wedge..\wedge
d\Phi^{n}\right|  $ is independent of $t$. If this integral is finite, it may
be assumed to be equal to $1$ without loss of generality. In this case each of
the hypersurfaces $\Sigma_{\Phi,t}$ carries the probability measure defined
for the Borel sets $X\subset\Sigma_{\Phi,t}$%
\[
M_{t}(X):=\int_{X}\left(  \psi^{\ast}\psi\right)  \cdot\left|  d\Phi^{1}
\wedge..\wedge d\Phi^{n}\right|  =\int_{X}\left|  J\right|  ,
\]
where $J=j\lrcorner\omega$ with $\omega$ chosen from $\left\{  \pm d\Phi
^{0}\wedge..\wedge d\Phi^{n}\right\}  $. The form $J$ is closed because of
$\div(j)=0$.

In case of $\psi^{\ast}\psi>0$ the vector field $\widehat{j}$ is defined on
all of $\mathcal{M}$. If $\widehat{j}$ is complete, its global flow $F$
provides a fibration of $\mathcal{M}$ by its orbits. The mappings $F_{t}$
evolve instantaneous regions from $\Sigma_{\Phi,s}$ into instantaneous regions
from $\Sigma_{\Phi,s+t}$ of the same probability content. Thus the orbit space
carries the unique probability measure, given by
\[
\mu(Y):=M_{t}\left(  \left\{  x\in\Sigma_{\Phi,t}\mid\exists o\in Y\mbox{ with
}x\in o\right\}  \right)
\]
for any $t\in\mathbb{R}$. Thus for the transition of $j$ through a set
$X\subset\mathcal{M}$ there holds $P\left[  X\right]  =\mu(\widetilde{X}
)\in\left[  0,1\right]$. Here $\widetilde{X}$ denotes the set of F-orbits intersecting
$X$.

Bohmian mechanics proposes to take serious the flow lines, i.e. the orbits of
$\widehat{j\mbox{,}}$ as the possible worldlines of a quantum point particle
with the wave function $\psi$. Which orbit is realised in each individual case
of an ensemble, is considered as being beyond experimental control, and is
assumed to be subject to the probability measure represented by $M_{0}$.
In this way Bohmian mechanis provides a picture of a world with facts,
evolving continuously in time, while simultaneously the quantum mechanical
expectation values of fixed time measurements remain unaltered. A
generalisation of Bohmian mechanics to wave functions, that do not yield a
globally defined complete velocity vector field, has been established in
\cite{BDG}.

Within the Bohmian extension of quantum mechanics, the following notion of
\textit{detection probability} seems plausible. The probability that the
Bohmian orbit of a (free) particle with wave function $\psi\in\ker
(D_{\underline{\partial}^{\Phi}})$ passes a given spacetime region
$X\subset\mathcal{M}$, equals the transition $P\left[  X\right]  $ of the
current vector field $j(\psi,\underline{\partial}^{\Phi})$ through $X$.
Observe that $P\left[  X\right]  $ does not depend on the choice of $\Phi
\in\mathcal{A}_{\mathcal{G}}$ and that indeed $0\leq P\left[  X\right]  \leq1$
holds. We now suggest that an (idealised) detector, which is sensitive to the
spacetime region $X$, registers the particle if and only if the particle's
Bohmian trajectory passes $X$. Therefore we assume the \textit{detection
probability within the spacetime region }$X$ to equal $P\left[  X\right]  $.

Let us consider a more specific situation. Let the set $X\subset\mathcal{M}$
be the union of time translates of a Borel subset $D$ of the instantaneous
space $\Sigma_{\Phi,0}$, i.e.
\[
\Phi(X):=\left\{  (t,x)^{t}\mid T_{1}\leq t\leq T_{2}\mbox{ and }(0,x)^{t}
\in\Phi\left(  D\right)  \right\}
\]
for given $T_{1}\leq T_{2}\in\mathbb{R}$. The set $X$ contains the spacetime
points covered by an inertial, rigid detector, which is activated at time
$T_{1}$ and which is turned off at time $T_{2}$. The number $P\left[
X\right]  $ is the probability that this detector clicks.

The mapping
\[
\delta:\left\{  (T_{1},T_{2})\in\mathbb{R\times R}\mid T_{1}\leq
T_{2}\right\}  \rightarrow\left[  0,1\right]  ,(T_{1},T_{2})\mapsto P\left[
X\right]
\]
is continuous. Furthermore the function $T_{2}\mapsto\delta(T_{1},T_{2})$ is
nondecreasing and the function $T_{1}\mapsto\delta(T_{1},T_{2})$ is
nonincreasing. Thus turning off later with $T_{1}$ being kept fixed does not
diminish and activating later with $T_{2}$ being kept fixed does not increase
the detection probablity.

In the next section we shall make use of the $\hbar=1$ and $m=1$
simplification of Schr\"{o}dinger's equation. This is obtained by introducing
the affine (non Galilean) chart $\chi=(\chi^{0},\chi^{1},..\chi^{n}
)=(\frac{1}{\hbar}\Phi^{0},\frac{\sqrt{m}}{\hbar}\Phi^{1},...\frac{\sqrt{m}
}{\hbar}\Phi^{n})$. Therefore we have
\begin{eqnarray*}
d\chi^{0}  & =\frac{1}{\hbar}d\Phi^{0},\quad d\chi^{1}=\frac{\sqrt{m}}{\hbar
}d\Phi^{1},...d\chi^{n}=\frac{\sqrt{m}}{\hbar}d\Phi^{n},\\
\partial_{0}^{\Phi}  & =\frac{1}{\hbar}\partial_{0}^{\chi},\quad\partial
_{1}^{\Phi}=\frac{\sqrt{m}}{\hbar}\partial_{1}^{\chi},..\partial_{n}^{\Phi
}=\frac{\sqrt{m}}{\hbar}\partial_{n}^{\chi}.
\end{eqnarray*}
Then $\psi\in\ker(D_{\underline{\partial}^{\Phi}})$ is equivalent to
\[
i\partial_{0}^{\chi}\psi=-\frac{1}{2}\sum_{k=1}^{n}\partial_{k}^{\chi}\left(
\partial_{k}^{\chi}\psi\right)  .
\]
The current vector field $j$, given by equation (\ref{stromVF}), and the
volume form $\omega:=d\Phi^{0}\wedge..\wedge d\Phi^{n}$ have the following
coordinate expressions in terms of $\chi$.
\begin{eqnarray*}
j  & =\frac{1}{\hbar}\left\{  \psi^{\ast}\psi\partial_{0}^{\chi}+\frac{1}
{2i}\sum_{k=1}^{n}\left[  \psi^{\ast}\left(  \partial_{i}^{\chi}\psi\right)
-cc\right]  \partial_{i}^{\chi}\right\}  ,\\
\omega & =\frac{\hbar^{n+1}}{m^{\frac{n}{2}}}d\chi^{0}\wedge..\wedge d\chi
^{n}.
\end{eqnarray*}
Thus in terms of the rescaled wave function $\Psi:=\left(  \frac{\hbar}
{\sqrt{m}}\right)  ^{\frac{n}{2}}\psi$ the current form $J=j\lrcorner\omega$
finally reads as follows
\[
J=\Psi^{\ast}\Psi d\chi^{1}\wedge...\wedge d\chi^{n}-\frac{1}{2i}\left[
\Psi^{\ast}\left(  \partial_{1}^{\chi}\Psi\right)  -cc\right]  d\chi^{0}\wedge
d\chi^{2}\wedge..\wedge d\chi^{n}+...
\]

\section{$P(T)$ for a Gaussian wave packet}

\subsection{The flow map}

We assume $n=1$ in what follows and we use the more suggestive notation:
$\chi^{0}=:\tau$ and $\chi^{1}=:\xi$. Accordingly we abbreviate: $\partial
_{0}^{\chi}=\partial_{\tau}$ and $\partial_{1}^{\chi}=\partial_{\xi}$. Let
$\delta\in\mathbb{R}_{>0}$. Then the complex valued function $\psi$ on
$\mathcal{M}$
\[
\psi:=\sqrt{\frac{\sqrt{m}}{\hbar}}\Psi\mbox{ with }\Psi:=\frac{1}
{\sqrt{\delta\sqrt{\pi}}}\cdot\frac{1}{\sqrt{1+i\frac{\tau}{\delta^{2}}}}
\cdot\exp\left[  -\frac{\xi^{2}}{2\delta^{2}}\cdot\frac{1}{1+i\frac{\tau
}{\delta^{2}}}\right]
\]
solves the Schr\"{o}dinger equation, i.e. $D_{\underline{\partial}^{\Phi}}
\psi=0$. It is a Gaussian wave packet centered at $\xi=0$ at all times. The
complex square root has its cut along the negative real axis. The current
vector field $j:=j(\psi,\underline{\partial}^{\Phi})$ is given by
\begin{eqnarray*}
j  &=& \frac{\sqrt{m}}{\hbar^{2}}\Psi^{\ast}\Psi\left[  \partial_{\tau
}+\frac{\tau\xi}{\delta^{2}\Delta^{2}}\partial_{\xi}\right]  ,\mbox{ with}\\
\Psi^{\ast}\Psi &=& \frac{1}{\sqrt{\pi}\Delta}\cdot\exp(-\frac{\xi^{2}}
{\Delta^{2}}).
\end{eqnarray*}
Here the positive realvalued function $\Delta$, defined on $\mathcal{M}$, is
given by
\[
\Delta:=\delta\sqrt{1+\left(  \frac{\tau}{\delta^{2}}\right)^{2}}.
\]
For later use we introduce the rescaled current $s:=\frac{\hbar^{2}}{\sqrt{m}
}j=s^{0}\partial_{\tau}+s^{1}\partial_{\xi}$. The velocity vector field
associated with $j$
\[
\widehat{j}=\partial_{\tau}+\frac{\tau\xi}{\delta^{2}\Delta^{2}}\partial_{\xi}
\]
is of $C^{\infty}$-type on $\mathcal{M}$.

The integral curves $\gamma_{p}$ of the velocity vector field $\widehat{j}$
through a point $p\in\mathcal{M}$ are obtained in terms of the functions
$x^{0}:=\tau\circ\gamma_{p}$ and $x^{1}:=\xi\circ\gamma_{p}$. They solve the
system of first order differential equations
\begin{eqnarray*}
\dot{x}^{0}  & =& 1,\\
\dot{x}^{1}  & =& \frac{x^{0}x^{1}}{\delta^{4}\left(  1+\left(  \frac{x^{0}
}{\delta^{2}}\right)  ^{2}\right)  }
\end{eqnarray*}
with the initial condition $p^{0}:=x^{0}(0)=\tau(p)$ and $p^{1}:=x^{1}
(0)=\xi(p)$. The first differential equation has the unique, maximal solution
$x^{0}(\lambda)=\lambda+p^{0}$ for any $\lambda\in\mathbb{R}$. Inserting this
solution into the second equation yields the non autonomous first order
differential equation
\[
\dot{x}^{1}(\lambda)=\frac{\left(  p^{0}+\lambda\right)  \cdot x^{1}(\lambda
)}{\delta^{4}\left(  1+\left(  \frac{p^{0}+\lambda}{\delta^{2}}\right)
^{2}\right)  }.
\]
Its unique, maximal solution is obtained by separation of variables. It is
given by
\[
x^{1}(\lambda)=p^{1}\sqrt{\frac{1+\left(  \frac{p^{0}+\lambda}{\delta^{2}
}\right)  ^{2}}{1+\left(  \frac{p^{0}}{\delta^{2}}\right)  ^{2}}}
\]
for any $\lambda\in\mathbb{R}$. Thus the vector field $\widehat{j}$ is
complete and the flow $F:\mathbb{R}\times\mathcal{M}\rightarrow\mathcal{M}$
defines a one parameter group of global diffeomorphisms $\left\{  F_{\lambda
}\mid\lambda\in\mathbb{R}\right\}  $ of $\mathcal{M}$. The coordinate
expression of $F_{\lambda}$ is as follows.
\[
\Phi\circ F_{\lambda}\circ\Phi^{-1}:\mathbb{R}^{2}\rightarrow\mathbb{R}
^{2},\left(  p^{0},p^{1}\right)  ^{t}\mapsto\left(  \lambda+p^{0},p^{1}
\sqrt{\frac{1+\left(  \frac{p^{0}+\lambda}{\delta^{2}}\right)  ^{2}}{1+\left(
\frac{p^{0}}{\delta^{2}}\right)  ^{2}}}\right)  ^{t}
\]

The (maximal) integral orbit of $\widehat{j}$ through $p\in\Sigma_{\Phi,0}$
is the set of points $\Gamma_{p}\subset\mathcal{M}$ on which holds $\xi
=p^{1}\frac{\Delta}{\delta}$. It is the well known hyperbolic worldline of
the Bohmian particle with wave function $\psi$ and passing through $p$. See
e.g. sect.4.7 of ref. \cite{Hol}. Some orbits are shown by figure 2 in terms
of the dimensionless coordinates $t:=\tau/\delta^{2}$ and $x:=\xi/\delta$.

\begin{figure}[ht]
\centering
\includegraphics[height=2.2018in,
width=2.1854in]{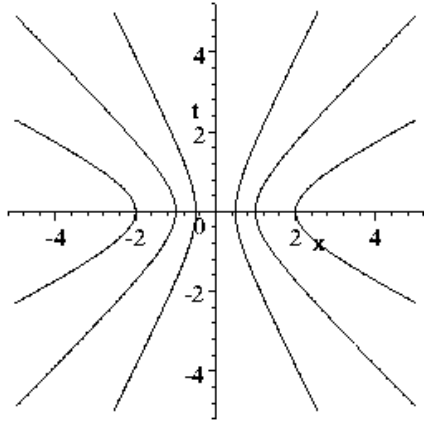}\\
\caption{Bohmian orbits of a Gaussian wave packet}
\end{figure}

\bigskip

The $1$-form $J=j\lrcorner\omega=s^{0}d\xi-s^{1}d\tau\;$obeys
\[
J=\Psi^{\ast}\Psi\left[  d\xi-\frac{\tau\xi}{\delta^{2}\Delta^{2}}
d\tau\right]  .
\]
Due to Poincare's lemma, $J$ is exact, i.e. there exist functions
$H:\mathcal{M}\rightarrow\mathbb{R}$ with $J=dH$. For any two functions
$H_{1}$ and $H_{2}$ with $dH_{1}=dH_{2}=J$ the difference $H_{1}-H_{2}$ is
constant on $\mathcal{M}$. Due to $dH=\left(  \partial_{\tau}H\right)
d\tau+\left(  \partial_{\xi}H\right)  d\xi$, for the function $H$ there holds
\begin{eqnarray*}
\partial_{\xi}H  &=& J(\partial_{\xi})=s^{0}=\Psi^{\ast}\Psi \quad\mbox{ and}\\
\partial_{\tau}H  &=& J(\partial_{\tau})=-s^{1}=-\frac{\tau\xi}{\delta
^{2}\Delta^{2}}\Psi^{\ast}\Psi \mbox{ .}
\end{eqnarray*}
A solution to these equations is given by
\[
H:=\frac{1}{2}\erf\left(  \frac{\xi}{\Delta}\right)  \mbox{ ,}
\]
where $\erf:\mathbb{R}\rightarrow(-1,1)$ denotes Gauss's error
function
\[
\erf(x):=\frac{2}{\sqrt{\pi}}\int_{0}^{x}\exp(-z^{2})dz \mbox{ .}
\]
Obviously, $H$ is constant on the orbits of $\widehat{j}$. This is due to
$dH(j)=J(j)=\omega(j,j)=0$.

\subsection{Detector activated at time $0$}

Now we shall discuss the detection probability of a pointlike detector, wich
is exposed to the wave function $\psi$. The detector is assumed to be located
at $\xi=L>0$ and is activated at $\tau=0$. Thus the detector measures the
transition of the current $j$ through the spacetime regions
\[
D_{T}:=\left\{  p\in\mathcal{M}\mid\xi(p)=L\mbox{ and }0\leq\tau(p)\leq
T\right\}  \quad\mbox{with\ } T>0.
\]
The boundary of $D_{T}$ equals $\left\{  A,B\right\}  $ with $(\tau
,\xi)(A)=(0,L)$ and $(\tau,\xi)(B)=(T,L)$ (see figure 3).

\begin{figure}[ht]
\centering
\includegraphics[height=2.1793in,
width=1.7573in]{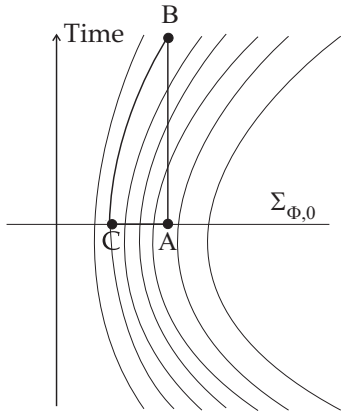}\\
\caption{Detector at rest from $A$ to $B$}
\end{figure}

The set of points $p_{0}\in\Sigma_{\Phi,0}$ whose integral orbits
$\Gamma_{p_{0}}$\ intersect $D_{T}$ is, due to $\Delta(p_{0})=\delta$,
\[
\pi\left(  D_{T}\right)  =\left\{  p_{0}\in\Sigma_{\Phi,0}\mid\mbox{there
exists a }p\in D_{T}\mbox{ with }\frac{\xi(p)}{\Delta(p)}=\frac{\xi(p_{0}
)}{\delta}\right\}  .
\]
Thus we obtain
\[
\pi\left(  D_{T}\right)  =\left\{  p_{0}\in\Sigma_{\Phi,0}\mid\frac{L\delta
}{\Delta(B)}\leq\xi(p_{0})\leq\frac{L\delta}{\Delta(A)}\right\}  .
\]
Due to $\Delta(A)=\delta$ and $\Delta(B)=\delta\sqrt{1+\left(  \frac{T}
{\delta}\right)  ^{2}}$, this yields
\[
\pi\left(  D_{T}\right)  =\left\{  p_{0}\in\Sigma_{\Phi,0}\mid\frac{L}
{\sqrt{1+\left(  \frac{T}{\delta}\right)  ^{2}}}\leq\xi(p_{0})\leq L\right\}
.
\]
The boundary of the line segment $\pi\left(  D_{T}\right)  $ equals $\left\{
A,C\right\}  $ with
\[
(\tau,\xi)(C)=(0,\frac{L}{\sqrt{1+\left(  \frac{T}{\delta}\right)  ^{2}}}).
\]
The detection probability $P\left[  D_{T}\right]  $ then follows by
integrating $\left|  J\right|  $ over $\pi(D_{T})$.
\begin{eqnarray*}
P\left[D_{T}\right] &=& \int_{\pi\left(D_{T}\right)} \left| J\right|
= \int_{\pi\left(D_{T}\right)}\left| \left(\partial_{\xi}H\right)
d\xi\right| = H(A)-H(C)\\
&=& \frac{1}{2}\left[  \erf\left(\frac{L}{\delta}\right)-\erf
\left(\frac{L}{\delta\sqrt{1+\left(  \frac{T}{\delta}\right)  ^{2}}}\right)\right]
=:\delta_{L}(0,T)
\end{eqnarray*}
The function $\delta_{L}(0,\cdot)$ is monotonically increasing, has the value
$0$ at $T=0$ and tends to $\frac{1}{2}\erf(\frac{L}{\delta}
)\in\left(  0,\frac{1}{2}\right)  $ for $T\rightarrow\infty$. The detection
probability stays below $1/2$ because no left moving orbit intersects with the
detection region $D_{T}$. The limit of a far away detector yields
$\lim_{L\rightarrow\infty}\lim_{T\rightarrow\infty}\delta_{L}(0,T)=1/2$.

Figure 4 shows $P\left[  D_{T}\right]  $ as a function of the dimensionless
time $t:=\frac{T}{\delta^{2}}$ for $L=100\delta$, i.e. the function
\[
f:R_{\geq0}\rightarrow\left[  0,1\right]  ,\quad t\mapsto\frac{1}{2}\left(
\erf(100)-\erf\Big(\frac{100}{\sqrt{1+t^{2}}}\Big)\right)
.
\]

\begin{figure}[ht]
\centering
\includegraphics[height=1.6207in,
width=3.1894in]{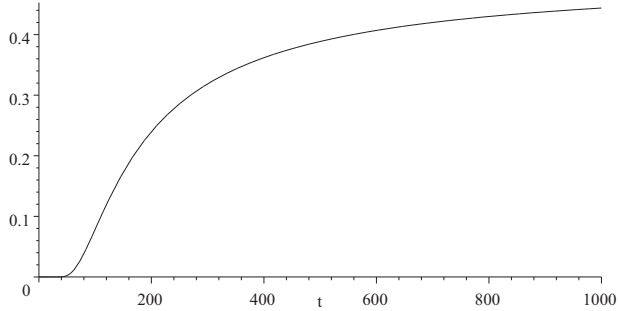}\\
\caption{Detction probability $P\left[  D_{T}\right]$}
\end{figure}

The equality between $P\left[  D_{T}\right]  $ and $P(T)$ as given by Leavens
\cite{L1993}, \cite{McL}, \cite{L1998}, we denote it $P_{L}(T)$, can be
derived as follows. The line segment $D_{T}$ has the boundary points $A$ and
$B$. The points $B$ and $C$ belong to the same orbit $\Gamma_{B}$ of $j$. The
part of $\Gamma_{B}$ lying inbetween $B$ and $C$ is denoted by $\Gamma_{B,C}$
Thus the union of the three segments $D_{T}$, $\Gamma_{B,C}$ and $\pi\left(
D_{T}\right)  $ is a closed line $K\subset\mathcal{M}$. The orientation of $K$
and its boundary $\partial K$ is determined by the chosen $\omega$
\cite{Stern}. Application of Stoke's theorem to the spacetime region $K$
interior to this closed line gives
\begin{eqnarray*}
0  &=& \int_{K}dJ\\
&=& \int_{\pi\left(  D_{T}\right)  }J+\int_{D_{T}}J+\int_{\Gamma_{B,C}}J.
\end{eqnarray*}
Since $\int_{\Gamma_{B,C}}J=0$, because of $\widehat{j}\lrcorner J=0$, and
$s^{0},s^{1}\geq0$ on $\partial K$, we obtain from this
\[
P\left[  D_{T}\right]  =\int_{\pi\left(  D_{T}\right)  }\left|  J\right|
=\int_{D_{T}}\left|  J\right|  =\int_{D_{T}}\left|  s^{1}d\tau\right|
=:P_{L}(T).
\]
Due to $H(B)=H(C)$, one explicitly verifies
\[ 
P_{L}(T)=\int_{D_{T}}\left| J\right| = \int_{D_{T}}\left| dH\right|
= \int_{D_{T}}\left| \left(\partial_{\tau}H\right) d\tau\right|
= H(A)-H(B) = P\left[D_{T}\right] .
\]
Thus in the present case the detection probability $P\left[D_{T}\right]$
is obtained by integrating the density $\left| s^{1}d\tau\right|$ along the
detector worldline $D_{T}$. Obviously, the equation
\begin{equation}
P\left[  D_{T}\right]  =\int_{D_{T}}\left|  s^{1}d\tau\right|
\label{Leavensdichte}
\end{equation}
is due to the absence of multiple intersections between $D_{T}$ and the
individual Bohmian orbits. We shall construct an explicit counterexample to
equation (\ref{Leavensdichte}) in the next subsection.

From the function $\delta_{L}(0,\cdot)$, the conditional probability density
of arrival times at a detector, which is activated at $\tau=0$, can be
obtained as follows. The conditioning is with respect to those events, where
the particle is detected at all by this detector. Define the normalised
conditional distribution function $W(T):=\frac{\delta(0,T)}{\lim
_{T\rightarrow\infty}\delta(0,T)}=\frac{H(A)-H(B)}{H(A)}$. The differential
$dW$ yields the conditional probability density $w\left|  dT\right|  :=\left|
dW\right|  $ of detection times. Thus $w(T)=\frac{dW(T)}{dT}$.
\begin{eqnarray*}
w(T)  &=& \frac{-\left(  \partial_{\tau}H\right)  (B)}{H(A)}=\frac{j^{1}%
(B)}{H(A)}\\
&=& \frac{1}{\erf\left(  \frac{L}{\delta}\right)  }\left(
-\partial_{\tau}\erf\Big(\frac{\xi}{\Delta}\Big)\right)  (B)\\
&=& \frac{2}{\sqrt{\pi}\erf\left(  \frac{L}{\delta}\right)
}\cdot\frac{LT}{\delta^{5}\left(  1+\left(  \frac{T}{\delta^{2}}\right)
^{2}\right)  ^{\frac{3}{2}}}\cdot\exp\left(  -\frac{L^{2}}{\delta^{2}\left(
1+\left(  \frac{T}{\delta^{2}}\right)  ^{2}\right)  }\right)  .
\end{eqnarray*}

The density $\widetilde{w}$ of the dimensionless time $t:=T/\delta^{2}$ is
defined through $\widetilde{w}(t)dt=w(T)dT$ and thus with $\lambda:=L/\delta$
we obtain
\[
\widetilde{w}(t)=\frac{2\lambda}{\sqrt{\pi}\erf(\lambda)}%
\cdot\frac{t}{\left(  1+t^{2}\right)  ^{\frac{3}{2}}}\cdot\exp\left(
-\frac{\lambda^{2}}{1+t^{2}}\right)  .
\]
Figure 5 shows the graph of $\widetilde{w}$ for $\lambda=100$.

\begin{figure}[ht]
\centering
\includegraphics[height=1.9899in,
width=2.2286in]{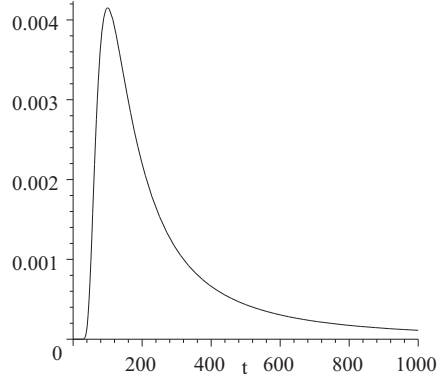}\\
\caption{Conditional probability density $\widetilde{w}$ of
arrival times}
\end{figure}

Since $\lim_{t\rightarrow\infty}t^{2}\widetilde{w}(t)>0$, the improper
integral $\lim_{\Lambda\rightarrow\infty}\int_{0}^{\Lambda}t\widetilde
{w}(t)dt$ does not exist. Thus an average (conditional) detection time does
not exist as well.

\subsection{Detector activated before time $0$}

In order to be sensitive to the contractive phase of the wave function, we now
assume that the detector is turned on at some time $T_{A}<0$. It thus measures
the transition through the sets of spacetime points
\[
D_{T}:=\left\{  p\in\mathcal{M}\mid\xi(p)=L\mbox{ and }T_{A}\leq\tau(p)\leq
T\right\}  \quad\mbox{with\ } T>T_{A}.
\]
The bounary $\partial D_{T}$ equals $\left\{  A,B\right\}  $, where $(\tau
,\xi)(A)=(T_{A},L)$ with $T_{A}<0$, $L>0$ and $(\tau,\xi)(B)=(T,L)$ (see
figure 6). We shall see the difference between $P\left[  D_{T}\right]  $ and
$P(T)$ according to Leavens \cite{L1998}, we again denote it as $P_{L}(T)$, clearly.

\begin{figure}[ht]
\centering
\includegraphics[height=2.4647in,
width=1.9865in]{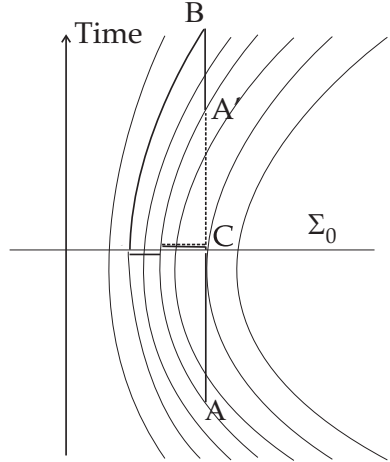}\\
\caption{Detector at rest from $A$ to $B$}
\end{figure}

The transition $P\left[  D_{T}\right]  $ then follows by inspection of
$\pi_{\Phi,0}(D_{T})$. With the auxilliary point $C:=D_{T}\cap\Sigma_{\Phi,0}$ we
obtain in terms of the dimensionless coordinates $t=T/\delta^{2},t_{A}
=T_{A}/\delta^{2},\lambda=L/\delta$
\begin{eqnarray*}
P\left[  D_{T}\right]   &=& \left\{
\begin{array}[c]{ccl}
H(B)-H(A) & \mbox{for} & T_{A}\leq T<0\\
H(C)-H(A) & \mbox{for} & 0\leq T<-T_{A}\\
H(C)-H(B) & \mbox{for} & -T_{A}\leq T
\end{array}
\right. \\
&=& \left\{
\begin{array}[c]{lcl}
\frac{1}{2}\left( \erf\left(\frac{\lambda}{\sqrt{1+t^{2}}}\right)
- \erf\Big(\frac{\lambda}{\sqrt{1+t_{A}^{2}}}\Big) \right)
& \mbox{for} & t_{A}\leq t<0\\
\frac{1}{2}\left(  \erf\left(\lambda\right)
-\erf\Big(\frac{\lambda}{\sqrt{1+t_{A}^{2}}}\Big) \right)
& \mbox{for} & 0\leq t<-t_{A}\\
\frac{1}{2}\left(  \erf\left(  \lambda\right)
-\erf\left(\frac{\lambda}{\sqrt{1+t^{2}}}\right) \right)
& \mbox{for} & -t_{A}\leq t
\end{array}
\right.  .
\end{eqnarray*}
Figure 7 shows $P\left[  D_{T}\right]  $ (solid line) as a function of $t$ for
$\lambda=100$ and $t_{A}=-\sqrt{3}\cdot100$. For $t>0$ our expression
$P\left[  D_{T}\right]  $ for the detection probability $P(T)$ differs
considerably from the integral of $\left|  J\right|  $ over $D_{T}$, proposed
by Leavens to represent $P(T)$. This latter integral yields
\begin{eqnarray*}
P_{L}(T) &:= & \int_{D_{T}}\left|  J\right|  =\left\{
\begin{array}
[c]{lcc}%
H(B)-H(A) & \mbox{for} & t<0\\
2H(C)-H(A)-H(B) & \mbox{for} & t\geq0
\end{array}
\right. \\
&  & =\left\{
\begin{array}[c]{lcl}
\frac{1}{2}\left(  \erf\left(\frac{\lambda}{\sqrt{1+t^{2}}}\right)
- \erf\Big(\frac{\lambda}{\sqrt{1+t_{A}^{2}}}\Big) \right)
& \mbox{for} & t_{A}\leq t<0\\
\erf(\lambda)-\frac{1}{2}\left(  \erf\Big(\frac{\lambda}{\sqrt{1+t_{A}^{2}}}\Big)
+ \erf\left(\frac{\lambda}{\sqrt{1+t^{2}}}\right)  \right)
& \mbox{for} & t\geq0
\end{array}
\right.
\end{eqnarray*}
Its dependence of $t$ is shown for $\lambda=100$ and $t_{A}=-\sqrt{3}\cdot
100$\ as a dashed line in figure7.

\begin{figure}[ht]
\centering
\includegraphics[height=1.8282in,
width=2.9369in]{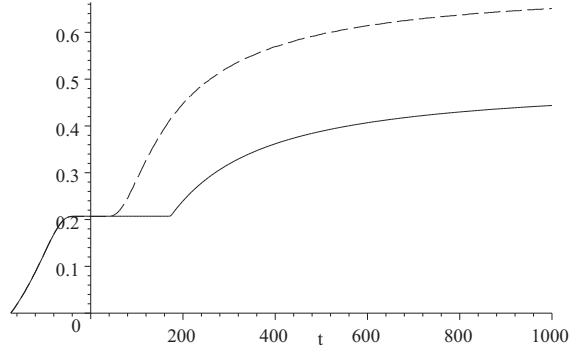}\\
\caption{Distribution functions $P\left[  D_{T}\right]  $ and
$P_{L}(T)$}
\end{figure}

$P\left[  D_{T}\right]  $ is constant for $0<T<-T_{A}$, while $P_{L}$ has a
point of stationarity only for $T=0$. For $0<T<-T_{A}$ orbits cross the
detector's worldline, which have done so before. Only past the point
$A^{\prime}$ with $\xi(A^{\prime})=L$ and $\tau(A^{\prime})=-T_{A}$ the
probability $P\left[  D_{T}\right]  $ increases again, because orbits are
passing, which have not done so before.

Figure 8 finally shows the conditional probabilities
\[
\frac{P(T)}{\lim_{T\rightarrow\infty}P(T)},
\]
associated with Leavens' proposal $P(T)=P_{L}(T)$ (dashed) and $P(T)=P\left[
D_{T}\right]  $ (solid) respectively.

\begin{figure}[ht]
\centering
\includegraphics[height=1.945in,
width=2.8876in]{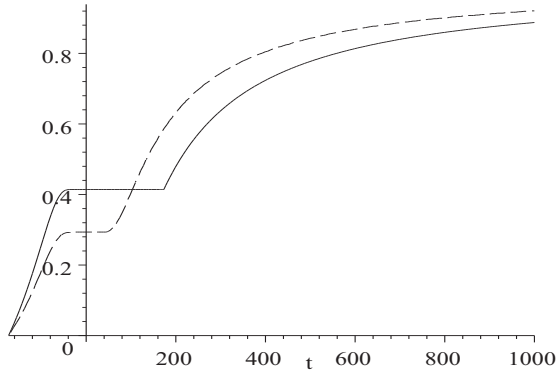}\\
\caption{Conditional distribution functions of $P_{L}(T)$ and
$P\left[ D_{T}\right]$}
\end{figure}

\section*{Acknowledgments} 
We are indebted to S Goldstein for a stimulating correspondence
and for bringing reference \cite{DDGZ2} to our attention. We thank H G
Embacher for \LaTeX\ support.

\end{document}